\documentclass[useAMS,usenatbib]{mn2e}
\usepackage{graphicx}

\newcommand{\spose}[1]{\hbox to 0pt{#1\hss}}
\newcommand{\approxpropto}{\mathrel{\spose{\lower 3pt\hbox{$\sim$}}
	\raise 2.0pt\hbox{$\propto$}}}
\def\approxgt{\mathrel{\spose{\lower 3pt\hbox{$\sim$}}
	\raise 2.0pt\hbox{$>$}}}
\def\approxlt{\mathrel{\spose{\lower 3pt\hbox{$\sim$}}
	\raise 2.0pt\hbox{$<$}}}

\def \hkpc{\,h^{-1} \hbox{kpc}}
\def \hMpc{\,h^{-1} \hbox{Mpc}}
\def \hMsol{\,h^{-1} \hbox{M}_{\odot}}

\def \h2gcm3{h^{2} \, \hbox{g}\,\hbox{cm}^{-3}}

\def \kevcm2{\, \hbox{keV} \, \hbox{cm}^{2}}
\def \cm3{\, \hbox{cm}^{-3}}

 \title[SZ temperature of the ICM]
       {The Sunyaev--Zel'dovich temperature of the intracluster medium}

\author[S.T. Kay et al.]
            {Scott T. Kay,$^{1}$\thanks{E-mail: Scott.Kay@manchester.ac.uk}
	      Leila C. Powell,$^{2}$
	     Andrew R. Liddle$^{3}$ and Peter A. Thomas$^{3}$\\ 
	$^{1}$Jodrell Bank Centre for Astrophysics, School of Physics and
	     Astronomy, The University of Manchester, Manchester M13 9PL\\
	$^{2}$Astrophysics, University of Oxford, Keble Road, 
              Oxford OX1 3RH, UK\\
	$^{3}$Astronomy Centre, Department of Physics and Astronomy,
            University of Sussex, Brighton BN1 9QH, UK\\ 
	}

\begin{document}

\date{This draft was generated on \today}

\pagerange{\pageref{firstpage}--\pageref{lastpage}} \pubyear{2005}

\maketitle

\label{firstpage}

\begin{abstract}
The relativistic Sunyaev-Zel'dovich (SZ) effect offers a method, 
independent of X-ray, for measuring the temperature of the intracluster 
medium (ICM) in the hottest systems. Here, using $N$-body/hydrodynamic 
simulations of three galaxy clusters, we compare the two quantities for a 
non-radiative ICM, and for one that is subject both to radiative 
cooling and strong energy feedback from galaxies. 
Our study has yielded two interesting results. Firstly, in all cases, 
the SZ temperature is hotter than the X-ray temperature and is within 
ten per cent of the virial temperature of the cluster. Secondly, the 
mean SZ temperature is less affected by cooling and feedback than the
X-ray temperature. Both these results can be explained by the SZ 
temperature being less sensitive to the distribution of cool gas 
associated with cluster substructure. A comparison of the SZ and X-ray 
temperatures (measured for a sample of hot clusters) would therefore yield 
interesting constraints on the thermodynamic structure of the
intracluster gas.
\end{abstract}


\section{Introduction}

Currently, nearly all of our knowledge of the intracluster medium
(ICM) comes from X-ray observations, with the latest generation of
X-ray satellites, {\it XMM-Newton} and {\it Chandra}, being capable of
spatially resolving the density and temperature structure
independently in low-redshift clusters out to tens of per cent of
their virial radii
(e.g. \citealt{Arnaud05,Vikhlinin05,Pratt07}). Separate density and
temperature information is vital for probing the effects of physical
processes (such as galactic outflows) on the structure of the ICM
(e.g. from the entropy and pressure profiles), as well as for
constructing mass--observable relationships for use in cosmological
analyses. The latter is currently being driven by X-ray 
surveys such as the {\it XMM} Cluster Survey \citep{Romer01,Stanford06}, 
promising to find many new clusters out to redshifts $z>1$.

The ICM is also detectable at centimetre and millimetre wavelengths
through the Sunyaev--Zel'dovich (SZ) effect \citep{SZ72}, the Inverse
Compton scattering of cosmic microwave background (CMB) photons off
free electrons in the ICM. As the SZ effect is a scattering process,
the observed SZ signal is independent of cluster redshift.  Thus,
thousands of new high-redshift clusters are promised from upcoming SZ
surveys such as that to be performed by the South Pole Telescope
\citep{Ruhl04}. 

For most clusters, the SZ effect will only measure the integrated
pressure of the electrons along the line of sight, and so its use as a
probe of ICM structure will be limited without additional X-ray
data. The combination of SZ and X-ray surface brightness profiles yields
a promising new way of obtaining cluster temperature information without
having to resort to expensive X-ray spectroscopy; such a method 
yields a temperature that is closer to the mass-weighted temperature 
of the cluster \citep{Ameglio07}. 

It is possible, however, to measure the temperature of the ICM
directly in the hottest clusters with multi-frequency SZ observations alone,
due to significant relativistic effects in these systems
(e.g. \citealt{Rephaeli95,Stebbins97,Itoh98,Challinor98,Pointecouteau98}).
In contrast to the above, this estimate of the temperature 
(hereafter refered to as the {\it SZ} temperature) is weighted 
by the pressure of the electrons along the line of sight.
How the SZ and X-ray temperature measurements differ systematically,
from a theoretical perspective, is the subject of this short
paper. Here, using $N$-body/hydrodynamic simulations of galaxy
clusters, we quantify the difference and show that it can be as large
as a factor of two, primarily due to the {\it clumpiness} in the ICM,
associated with incomplete thermal support. We also show that the 
temperature directly derived from SZ data is the more faithful tracer 
of the underlying cluster potential, being within 10 per cent of the 
virial temperature in all clusters studied.

\section{Measuring cluster temperatures}

The temperature of the ICM (or any portion of it) is most easily
measured by fitting a single-temperature plasma model to the observed
X-ray spectrum. As the emission is primarily due to thermal
bremsstrahlung (free--free) in hot clusters ($kT>2$ keV or so), the
temperature is determined by the energy scale, $E \simeq kT$, above
which the emission decays exponentially.  Since a single-temperature
model is used to fit a multi-temperature distribution, the result is
known as a spectroscopically-weighted average of the plasma
temperature.

Recently, \citet{Mazzotta04} calibrated this weighting using 
$N$-body/hydrodynamic simulations of galaxy clusters and proposed a simple 
estimator, known as the {\it spectroscopic-like} temperature
\begin{equation}
T_{\rm sl} = { \int \rho^2 T^{1/4}\, dV \over \int \rho^2 T^{-3/4} \, dV},
\label{eqn:tsl}
\end{equation}
where $\rho$ is the mass density of a fluid element in volume, $dV$,
with temperature, $T$. Since the weighting factor is $w=\rho^2 T^{-3/4}$,
cooler dense (i.e. lower entropy) gas is weighted the most. Naturally, 
cooling gas increases its density to try and re-establish hydrostatic 
equilibrium, so regions where cooling is important (particularly the core)
have a strong influence on the X-ray temperature of the ICM.

An alternative, but much harder, measurement of the ICM temperature comes
from the spectral distortion induced by a cluster on the CMB, the SZ 
effect. The fractional change in CMB temperature is
\begin{equation}
{\Delta T \over T_{\rm CMB}}  = 
\int \, \left[ g(x) {kT_{\rm e} \over m_{\rm e} c^2} +
               f(x,T_{\rm e}) {kT_{\rm e} \over m_{\rm e} c^2} +
               h(x) {v_{\rm l} \over c} \right] \, d\tau,
\label{eqn:szdef}
\end{equation}
where $T_{\rm CMB}=2.725$K \citep{Mather99}, $T_{\rm e}$ is the
electron temperature of the ICM, $n_{\rm e}$ its density, $v_{\rm l}$
the line-of-sight velocity, and $d\tau = n_{\rm e} \sigma_{\rm T} dl$
the change in optical depth to Compton scattering along the
differential line element, $dl$. The three frequency-dependent
pre-factors, $g(x),f(x,T_{\rm e}),h(x)$, where $x=h\nu/kT_{\rm CMB}$,
allow the separation of the SZ signal to be performed 
(with $f(x,0)=0$ for all $x$).

The first term in equation~(\ref{eqn:szdef}) dominates for reasonable
assumptions of cluster properties, and is due to the thermal energy of
the (mainly) non-relativistic electron population. The second term is
the temperature-dependent correction due to relativistic electrons
which can be significant for hot clusters; it is this term that we
will make use of to estimate the ICM temperature.  Finally, the third
term (known as the kinetic SZ effect) is due to the bulk motion of the
plasma.

The frequency-independent part of the first term is usually expressed
as the Compton $y$ parameter
\begin{equation}
y = \int \, {kT_{\rm e} \over m_{\rm e} c^2} \, d\tau.
\end{equation}
Following Hansen (2004), we can then define the Compton-average
of any quantity $F$ as
\begin{equation}
\left< F \right> = {1 \over y} \,
\int \, F \, {kT_{\rm e} \over m_{\rm e} c^2} \, d\tau,
\end{equation}
so the Compton-averaged electron temperature is
\begin{equation}
\left< T_{\rm e} \right> = 
{1 \over y} \,
\int \, T_{\rm e} \, 
     {kT_{\rm e} \over m_{\rm e} c^2} \, d\tau.
\label{eqn:tsz}
\end{equation}
The relativistic correction, $f(x,T_{\rm e})$, can be approximated as
$f(x,T_{\rm e})=T_{\rm e}\delta(x)$; higher-order corrections due to
ultrarelativistic electrons are subdominant in all but the very
hottest clusters
\citep{Diego03,Hansen04}.  In the Rayleigh--Jeans limit ($x \ll 1$),
$g(x)=-2$ and $\delta(x)=17k/5m_{\rm e}c^2$ \citep{Challinor98}.  Also
ignoring the kinetic SZ effect, we can thus approximate
equation~(\ref{eqn:szdef}) as
\begin{equation}
{\Delta T \over T_{\rm CMB}} = 
y [ g(x) + \delta(x) \left< T_{\rm e} \right> ].
\end{equation}
\begin{table}
\caption{Weights used to define the hot ($T>10^5$K) gas mass-weighted
temperature, $T_{\rm mw}$, the X-ray spectroscopic-like temperature,
$T_{\rm sl}$, and the SZ temperature, $T_{\rm SZ}$.}
\begin{center}
\begin{tabular}{ll}
\hline
Temperature & Weight\\
\hline
$T_{\rm mw}$ & $w=\rho$\\
$T_{\rm sl}$ & $w=\rho^2 T^{-3/4}$\\
$T_{\rm SZ}$ & $w=\rho T$\\
\hline
\end{tabular}
\end{center}
\label{tab:weights}
\end{table}
One can potentially extract $\left< T_{\rm e} \right>$ from
multi-frequency SZ data, especially for hot $kT>5$keV clusters.  This
is a different average from the X-ray temperature, weighted instead by
the gas pressure (so at a given density, the hottest, rather than
coolest, gas is weighted highest).  For simplicity, we will denote
this temperature $T_{\rm SZ}$ in subsequent discussion. A summary of
the weights used to define the gas temperatures is given in
Table~\ref{tab:weights}.

\section{Cluster Simulations}

We analyse $N$-body/hydrodynamic simulations of three galaxy clusters,
selected from a larger sample already studied by Kay et al.\ (2004;
hereafter K2004), to which we refer the reader for further
details. The clusters were extracted from a large cosmological
$N$-body simulation run by the Virgo Consortium,\footnote{{\tt
http://www.virgo.dur.ac.uk}} and re-simulated at higher resolution and
with gas using the {\sc gadget2} code \citep{Springel06}. The clusters
have similar masses [$M_{\rm vir}=(1.0,0.6,0.9)\times 10^{15} \hMsol$]
and radii [$R_{\rm vir}=(2.0,1.7,2.0) \hMpc$] at $z=0$, but were
chosen because they have significantly different merger histories
(Powell et al., in preparation). While this sample is limited in size
(primarily due to the amount of CPU time required), it nevertheless
gives some indication of object-to-object variations on the scale of a
typical rich cluster.
Each cluster was re-simulated twice (see below), with 
$1-2\times 10^6$ dark matter particles within $R_{\rm vir}$
(the dark matter and gas particle masses were 
$m_{\rm dark}=4\times10^{8}\hMsol$ and $m_{\rm gas}=8\times 10^{7}\hMsol$
respectively). 
The force
resolution (equivalent Plummer softening length) was fixed at 
$\epsilon=10\hkpc$ in comoving co-ordinates until $z=1$, after which 
it was fixed at $\epsilon=5\hkpc$ in physical co-ordinates.

\subsection{Cluster models}

For each cluster, we considered two models for the gas physics, taking
the total number of simulations performed to six. These models are
very similar to those studied by K2004, so further details may be
found there.

For the first model, labelled {\it non-radiative}, the gas was
subjected to adiabatic forces and an artificial viscosity (to generate
entropy in shocks) only. This was done using the standard SPH
implementation in {\sc gadget2} \citep{Springel06}. This model
contains the minimum amount of physics required to model the formation
of the ICM, being driven solely by gravitational processes.  While a
useful baseline, it does not match the observational properties of
clusters.

For the second, more realistic, model, labelled {\it feedback}, the
gas was subjected to the following additional processes. Firstly, gas
particles with temperature $T > 10^4$K were able to cool radiatively,
assuming a metallicity of $Z=0.3Z_{\odot}$.  Second, gas that had
cooled below $1.2\times 10^4$K and reached hydrogen densities, $n_{\rm
H}>10^{-3}{\rm cm}^{-3}$, could either form stars (i.e. become
collisionless) or be reheated to high temperature ($kT=25$keV), with
equal probability (i.e.  $f_{\rm heat}=0.5$ in the jargon of
K2004). Tests revealed that the choice of density threshold and
reheating temperature were necessary for the clusters to have
sufficiently-high core entropy to lie on the $z=0$ X-ray
luminosity--temperature relation, while the choice of $f_{\rm heat}$
was primarily to produce a sensible cooled fraction (on average, only
14 per cent of the baryons within $R_{500}$ had cooled by $z=0$,
similar to that observed by \citealt{Lin03}).
  
\subsection{Estimating the temperatures from the simulations}

The various 2D temperature distributions were computed from maps using
the procedure outlined in \citet{Onuora03}. Briefly, the relevant
weighting was computed for every hot ($T>10^5$K) gas particle within a
cylinder of length $6 R_{\rm vir}$, and projected radius $R_{\rm
vir}$, centred on the cluster.  These weightings were then smoothed,
and projected along the length of the cylinder onto a 2D $800 \times
800$ pixel array 
using the projected version of the {\sc gadget2} SPH kernel.  Maps of
the spectroscopic-like temperature, $T_{\rm sl}$, were computed using
the discrete version of equation~(\ref{eqn:tsl}) and the SZ
temperature, $T_{\rm SZ}$, using equation~(\ref{eqn:tsz}). For
comparison, we also computed hot gas mass-weighted temperature maps
(where the weight in equation~(\ref{eqn:tsl}) is replaced by $w=\rho$;
see Table~\ref{tab:weights}).

\section{Results}

\begin{table}
\caption{Various temperature measurements (expressed as an energy,
$kT$ in keV units) within the virial radius. Column~1 gives the 
label for each simulation; column 2 the virial temperature; 
column 3 the hot gas mass-weighted temperature; column 4 the 
X-ray spectroscopic-like temperature; column 5 the SZ 
Compton-weighted temperature and column 6 the ratio of the SZ
to X-ray temperatures. Values in brackets denote the ratio
of each temperature to the virial temperature.}
\begin{center}
\begin{tabular}{lccccc}
\hline
Cluster & $kT_{\rm vir}$ & $kT_{\rm mw}$ & $kT_{\rm sl}$ & $kT_{\rm
  SZ}$ & $T_{\rm SZ}/T_{\rm sl}$\\ 
\hline
NR1 & 5.7 & 4.2 (0.7) & 4.0 (0.7) & 6.1 (1.1) & 1.5\\
NR2 & 4.0 & 2.7 (0.7) & 2.2 (0.6) & 3.9 (1.0) & 1.8\\
NR3 & 6.0 & 3.8 (0.6) & 2.7 (0.5) & 5.6 (0.9) & 2.1\\
\hline
FB1 & 5.3 & 4.4 (0.8) & 5.4 (1.0) & 5.9 (1.1) & 1.1\\
FB2 & 3.9 & 3.3 (0.9) & 3.5 (0.9) & 4.4 (1.1) & 1.2\\
FB3 & 6.2 & 4.5 (0.7) & 4.8 (0.8) & 6.2 (1.0) & 1.3\\
\hline
\end{tabular}
\end{center}
\label{tab:temp}
\end{table}

We first present various measurements of the mean temperature within
the (projected) virial radius of each simulated cluster. Besides the
X-ray and SZ temperatures, we also consider the projected hot gas
mass-weighted temperature, $T_{\rm mw}$, and the 3D virial
temperature, $T_{\rm vir}$ (see \citealt{Muanwong02}), the most
suitable temperature for cosmological applications of clusters. 

The measured temperatures\footnote{In practical terms, the
temperatures are at the lower end of what might be measurable through
the SZ effect, but the overall results of this paper should not be
significantly affected by using hotter clusters.}  are given in
Table~\ref{tab:temp}.  The hot gas mass-weighted temperature, $T_{\rm
mw}$, and X-ray spectroscopic-like temperature, $T_{\rm sl}$, are both
lower than the virial temperature, with the exception of one of the
{\it feedback} clusters (FB1), where $T_{\rm sl} \simeq T_{\rm vir}$.  In
general, the difference between these two temperatures and the virial
temperature is larger for the {\it non-radiative} clusters than the
{\it feedback} clusters, the most extreme case being NR3, where
$T_{\rm sl} \simeq T_{\rm vir}/2$.  Part of this difference can be
attributed to projection effects, i.e. including gas extending to 3
virial radii from the cluster centre, along the line of sight, when
calculating $T_{\rm mw}$ and $T_{\rm sl}$. Cooler gas, associated with
infalling substructure, is more prevalent on the outskirts of
clusters, causing a significant downward shift in the measured
temperatures.  When the calculation of projected temperatures is
restricted to gas within $R_{\rm vir}$ in all directions, the
difference between these temperatures and $T_{\rm vir}$ is
approximately halved.

\begin{figure}
\centering
\includegraphics[width=8.5cm]{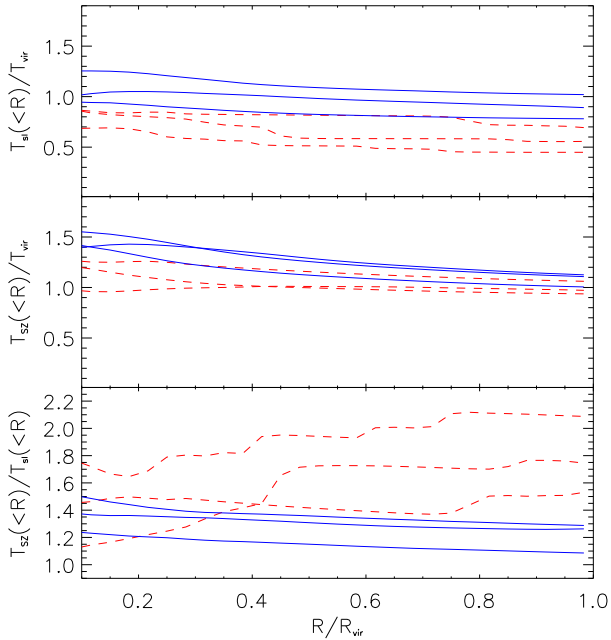}
\caption{Mean temperature within a projected radius, $R/R_{\rm vir}$,
for the {\it non-radiative} (dashed lines) and {\it feedback} (solid
lines) clusters. The top panel shows the (X-ray) spectroscopic-like
temperature profiles, the middle panel the SZ temperature profiles
(both panels show temperatures in relative to the virial temperature,
$T_{\rm vir}$) and the bottom panel the ratio of the two
temperatures.}
\label{fig:meant}
\end{figure}

We also show, in Fig.~\ref{fig:meant} (top panel), how $T_{\rm sl}$
compares to $T_{\rm vir}$ when the former is measured within a
projected radius $R<R_{\rm vir}$ (as is the case with X-ray
observations of clusters). Thus, the temperature at each value of $R$ is
a result of integrating over all gas within that radius.
At $R_{500}$ ($\sim R_{\rm vir}/2$), the
typical outer radius where $T_{\rm sl}$ can be reliably measured with
high-quality X-ray data, $T_{\rm sl}$ is only slightly larger (within
10 per cent) than its value at $R_{\rm vir}$. On the other hand, the
SZ temperature, $T_{\rm SZ}$, is higher than both $T_{\rm mw}$ and
$T_{\rm sl}$ in all clusters studied and is much closer to $T_{\rm
vir}$ (within 10 per cent).  All clusters show that $T_{\rm SZ}$
converges to within 20 per cent of $T_{\rm vir}$ at $R>0.6R_{\rm vir}$
($\sim R_{200}$). Finally, the ratio, $T_{\rm SZ}/T_{\rm sl}$ exhibits
very different behaviour between the two models, but at small radius
in particular, the SZ temperature can be significantly (tens of per
cent) higher than the X-ray temperature. At the virial radius, one
{\it non-radiative} cluster even has $T_{\rm SZ}>2T_{\rm sl}$, a
massive difference.

\begin{figure}
\centering \includegraphics[width=8.5cm]{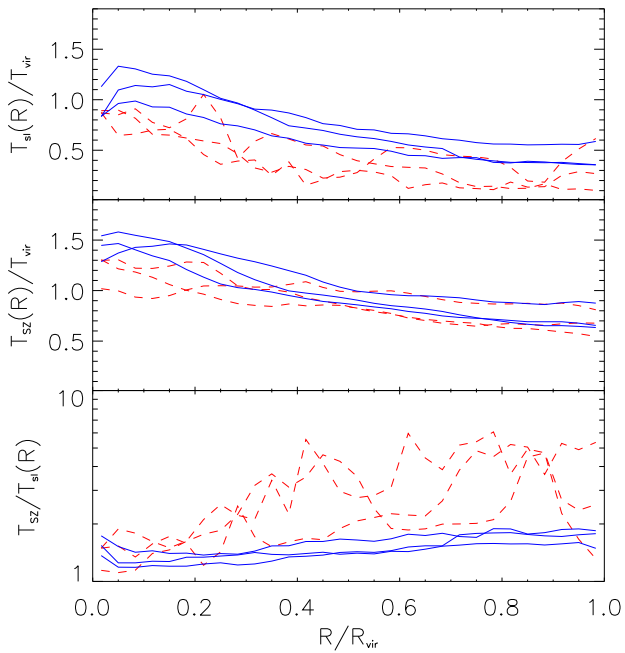}
\caption{Azimuthally-averaged temperature profiles for the {\it non-radiative}
(dashed lines) and {\it feedback} (solid lines) clusters.
The top panel shows the (X-ray) spectroscopic-like temperature profiles, 
the middle panel the SZ temperature profiles and the bottom panel the 
ratio of the two temperatures.}
\label{fig:profs}
\end{figure}

Figure \ref{fig:profs} illustrates how the two temperatures ($T_{\rm
sl}$ and $T_{\rm SZ}$) vary when measured locally,
azimuthally-averaged at each radius. Strikingly, the SZ temperature
profiles are very similar for both models outside $\sim 0.4 R_{\rm
vir}$, suggesting that this estimator is insensitive to
cluster physics away from the core. When the X-ray estimator is used,
the {\it feedback} clusters are systematically hotter than the {\it
non-radiative} clusters at all radii (except in the inner core, where
a sharp drop occurs due to the presence of gas that has radiated a
significant amount of its thermal energy).  The bottom panel in the
figure explicitly illustrates the ratio, $T_{\rm SZ}/T_{\rm sl}$, as a
function of radius. In both models, the ratio increases with radius,
although the effect is milder in the {\it feedback} clusters, with
$T_{\rm SZ}$ almost a factor of two higher than $T_{\rm sl}$ at $R_{\rm vir}$.


\begin{figure*}
\centering
\includegraphics[width=17cm]{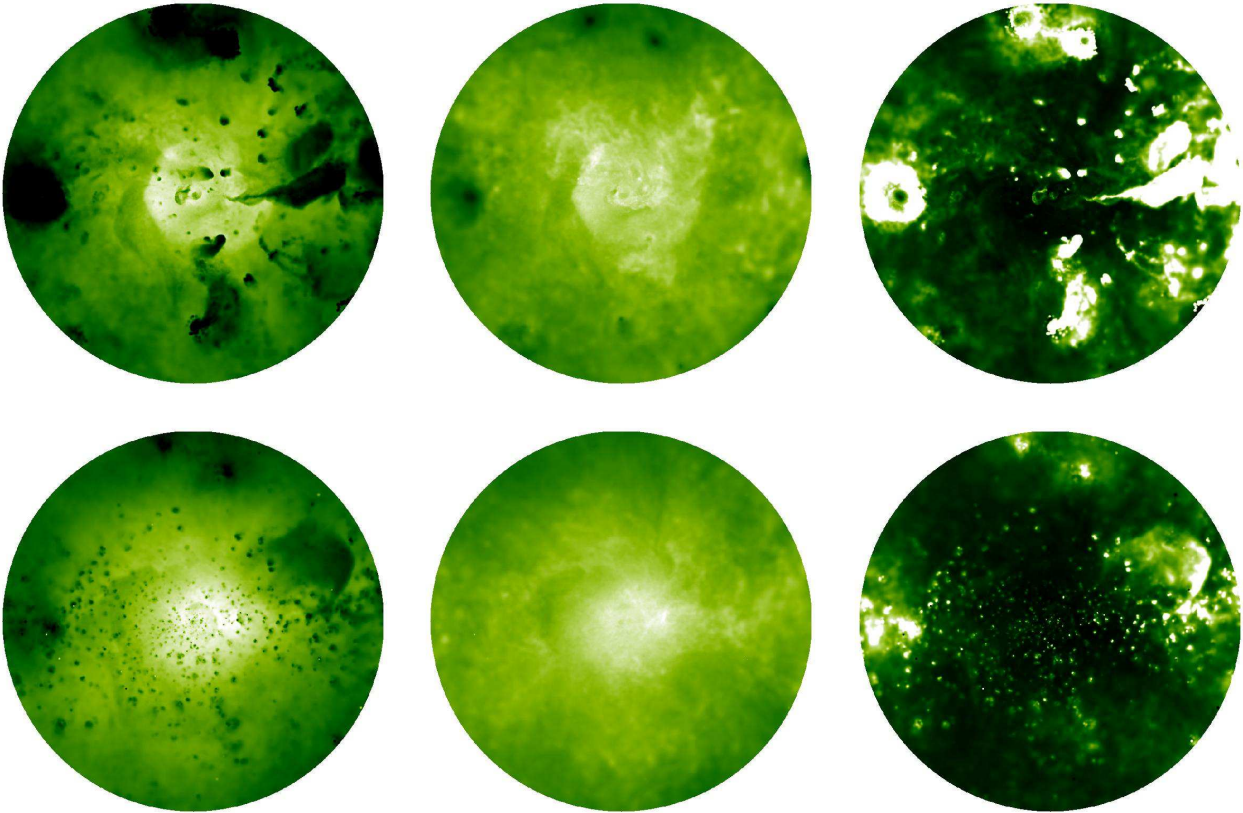}
\caption{Maps of the {\it non-radiative} 
(top panels) and {\it feedback} (bottom panels) versions of 
cluster 1 within their virial radius ($2\hMpc$). From left to right: 
X-ray spectroscopic-like temperature, Compton-weighted SZ temperature 
and the ratio of the two. For the first two maps, the colour scale 
illustrates logarithmic temperature values, $\log(T/{\rm K})=[7,8]$, 
while for the third, the scale illustrates ratios from 0 to 3.}
\label{fig:maps}
\end{figure*}

\begin{figure}
\centering
\includegraphics[width=8.5cm]{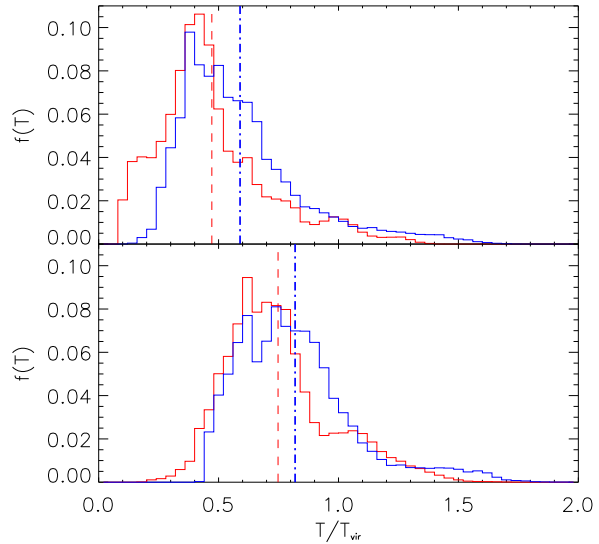}
\caption{X-ray (top panel) and SZ (bottom panel) temperature
distributions (fraction of pixels with a given $T/T_{\rm vir}$
value, within $R_{\rm vir}$)
for the {\it non-radiative} (thin lines) and {\it
feedback} (thick lines) versions of cluster 1. Vertical
dashed/dot-dashed lines illustrate the mean pixel value for the {\it
non-radiative}/{\it feedback} models.}
\label{fig:tdist}
\end{figure}

The reason why the SZ temperature is hotter than the X-ray temperature
(and closer to the virial temperature) can be seen in
Fig.~\ref{fig:maps}, where we present maps of the two quantities (and
their ratio) within $R_{\rm vir}$ for the first cluster (NR1 and FB1).
In the {\it non-radiative} cluster, the map contains a significant
amount of dark features, associated with infalling cooler, denser
gas. Since $T_{\rm sl}$ weights this gas higher, they make a prominent
contribution to the temperature of the cluster.  Note that the nature
of the cool gas changes in the {\it feedback} cluster. There it
becomes more spot-like, associated with the cores of subhaloes; the
combination of cooling and strong feedback has removed most of the
cool, low-entropy, gas in this model.  The difference in the amount of
cool gas in the X-ray and SZ temperature maps is shown clearly in
Fig.~\ref{fig:tdist}, where we plot pixel temperature distribution
functions within $R_{\rm vir}$. For the SZ distributions, the
differences at the lowest temperatures are also much less significant.

\section{Summary}

Our results clearly show that the temperature of the ICM, as measured
through the SZ effect, can differ significantly from the X-ray
temperature of the cluster, with the SZ temperature always being the
larger of the two. The X-ray temperature is weighted by cool dense gas
the most, so differences between the amount of cool {\it clumpy} gas
due to, for example, the level of radiative cooling and feedback, are
most significant. On the other hand, the SZ temperature weights the
hottest, densest gas (i.e. gas with the highest pressure) the most, so
is generally a better estimator of the virial temperature of the
cluster.  Consequently, differences due to the amount of cooling and
heating are less severe. The lower sensitivity of the SZ temperature to 
cluster physics (as compared to X-ray) is in line with previous comparisons 
between X-ray and SZ intensity (\citealt{daSilva01,daSilva04,Nagai06,Bonaldi07}).

In practice, the SZ temperature of a cluster is extremely difficult to
measure, and its applicability limited to the hottest clusters,
typically $kT>5$ keV. It does seem feasible, however, to design
multi-wavelength experiments that are able to constrain SZ
temperatures to $\sim 10-20$ per cent accuracy in the near future
\citep{Knox04}.  Whether an instrument will ever be able to measure
the SZ temperature in enough clusters, and to sufficient accuracy, to
be useful for cosmological purposes (where the virial temperature is
the desired quantity, c.f. \citealt{Evrard07}) remains to be seen, but
comparison with the X-ray temperature of a cluster ought to provide
useful information on the structure of the ICM.

\section*{Acknowledgements}
We would like to thank the anonymous referee for their report that
led to an improved version of this manuscript. We also thank 
Adrian Jenkins for generating the initial conditions
for the simulations used in this paper and Mike Jones for useful
discussions.

\bsp

\label{lastpage}

\end{document}